# Online and Utility-Power Efficient Task Scheduling in Homogeneous Fog Networks

Fatemeh Ebadi ; Vahid Shah-Mansouri

*Abstract*—Fog computing is of particular interest to Internet of Things (IoT), where inexpensive simple devices can offload their computation tasks to nearby Fog Nodes. Online scheduling in such fog networks is challenging due to stochastic network states such as task arrivals, wireless channels and location of nodes. In this paper, we focus on the problem of optimizing computation offloading management, arrival data admission control and resource scheduling, in order to improve the overall system performance, in terms of throughput fairness, power efficiency, and average mean of queue backlogs. We investigate this problem for a fog network with homogeneous mobile Fog Nodes, serving multiple wireless devices, controlled by a Fog Control Node. By formulating the problem as a stochastic optimization problem, maximizing utility-power efficiency, defined as achievable utility per-unit power consumption, subject to queue backlog stability, we modify Lyapunov optimization techniques to deal with the fractional form of utility-power efficiency function. Then we propose an online utility-power efficient task scheduling algorithm, which is asymptotically optimal. Our online task scheduling algorithm can achieve the theoretical $[O(1/V), O(V)]$ trade-off between utility-power efficiency and average mean of queue backlogs.

## I. Introduction

In recent years, intelligent mobile devices with wireless technologies have been widely used in industry and the Internet of Things (IoT). These wireless devices (WDs), which are incapable of handling massive computational tasks, due to their poor resources, e.g., the processing rate, memory and cache size, and battery capacity, should transfer the enormous data received from the ambient environment to more powerful remote machines to perform computation tasks. These computation tasks, like facial and gesture recognition, online 3D modeling, interactive games, and voice control, require a more stringent quality of computation experience. To do this, WD tasks will be offloaded and done at nearby Fog Nodes with spare computing, communication, and storage resources. Fog computing can relieve the tension between computation-intensive tasks and the inability of mobile WDs [1]–[3]. Contrary to conventional cloud computing systems, which rely on remote public cloud servers and result in increased delay, fog computing provides computation capability within the coverage area of a fog network. By offloading computation tasks from the WDs to the nearby Fog Nodes, the quality of computation experience, including power consumption, execution delay, and fairness among WDs, can be greatly improved. It is imperative to take into account the fairness of throughput by regulating the task arrival rates of each WD. The concept of enhancing service quality for computationally incapable devices, without any capital investment, solely through efficient management of the network of available resources, has attracted the attention of both industry and research [4].

To move towards practical implementations, optimal offloading schedules should be studied in depth, considering all the stochastics and variances of the system [5].

Numerous studies on fog computing have assumed predictability or neglected the stochasticity, and have examined deterministic optimization problems for resource scheduling and computation offloading, assuming complete knowledge of network states [3], [6]–[8]. However, many of the statistics utilized in their works are typically unknown in prior, and many of the proposed programming problems lack scalability due to inherent complexity issues.

Recently, a few studies have considered the stochasticity and unpredictability and have optimized resource scheduling foresightedly by using stochastic optimization techniques, i.e., Lyapunov optimization [9]–[14]. In [15], joint energy efficiency and application throughput fairness has been considered for mobile device resource management problem in computation offloading environment. However, these works have not jointly considered throughput fairness among devices and power consumption of the network subject to queue backlog stability in a fog network.

In this paper, we investigate the utility and power efficient task scheduling in a homogeneous fog network with a finite delay where the utility is defined as throughput fairness among WDs. In this case, computation and communication resources of Fog Nodes are dynamically and beneficially shared among WDs via the assistance of a Fog Control Node, as shown in Fig. 1. In our stochastic system, where the Fog Nodes and WDs are mobile and WDs requirements are dynamic, and the channel conditions also change over time, the Fog Control Node should design an online task scheduling algorithm. The algorithm should ensure utility-power efficiency in addition to maintaining the queue backlog stability running on each time slot. Dynamic computation offloading, CPU clock rate scheduling for each Fog Node, transmission power allocation, and data admission control in each WD are determined by the Fog Control Node according to the current network states and the arrival of WDs tasks. The key contributions of this paper are as follows:

- For a fog network with homogeneous Fog Nodes, with limited computation capability, serving multiple WDs via the assistance of a Fog Control Node, a collaborative utility-power efficient task scheduling problem subject to queue backlog stability is formulated. In this problem, we consider not only the power efficiency but also throughput fairness among WDs.

- A novel and online utility-power efficient task scheduling algorithm, revealing the trade-off between utility-power efficiency and the average mean of queue backlogs in the fog network, is proposed. Specifically, our algorithm modifies Lyapunov optimization techniques to maximize the overall utility-power efficiency while reducing the average mean of queue backlogs.

- Theoretical analysis of the trade-off between performance metrics, defined as utility-power efficiency and average mean of queue backlogs, is conducted for the proposed algorithm. And simulation results corroborate the theoretical analysis and show that the proposed algorithm can balance the utility-power efficiency and average mean of queue backlogs performance.

The rest of the paper is organized as follows. The system model is presented in Section II. The performance metrics and utility-power efficiency maximization problem is formulated in Section III. Online and utility-power efficient task scheduling algorithm is developed in Section IV. Long-term theoretical performance analysis is conducted in Section V. and evaluations of the proposed algorithm are shown in Section VI. Section VII concludes the paper.

## II. SYSTEM MODEL

As depicted in Fig. 1, we consider a fog network comprising a collection of $M = \{1, \ldots, |M|\}$ homogeneous mobile Fog Nodes, which execute computationally demanding tasks that are transferred via wireless links from heterogeneous low-mobility WDs indexed by $N = \{1, \ldots, |N|\}$, with the aid of a Fog Control Node. Similar scenarios exist, wherein computationally limited mobile devices offloads their computation tasks to Fog Nodes, such as the wireless sensor networks (WSNs) for surveillance and the IoT applications [16], [17]. The Fog Control Node has the potential to be installed in telecom infrastructures that are already in place. Therefore, it can be accessed by the Fog Nodes and WDs through wireless channels and determine computation offloading management, arrival data admission control, and resource scheduling with willingly forthcoming wireless channel state information at the Fog Control Node. To model our control over hardware power consumption and computation capacity, we assume that the Fog Node supports a variable clock rate. This is because a Fog Node may offer a faster service, resulting in increased processing capability, but at the cost of increased power consumption. Computation and communication resources of Fog Nodes are shared among WDs, via the Fog Control Node schedules. By setting up communication links with close by available and low-load Fog Nodes and offloading their computation tasks to them, WDs can opportunistically exploit the under-utilized computation resources of Fog Nodes. This will improve their quality of experience (QoE) and reduce their energy consumption [11].

The system operates in a slotted structure, $t \in T = \{0, 1, \ldots\}$ with slot length $\tau$. The Fog Control Node has access to global network information, including the location of Fog Nodes and WDs, computation resources of Fog Nodes, and task arrivals of WDs. Therefore, we consider a network-assisted architecture where the Fog Control Node coordinates task scheduling, resource discovery, arrival data admission, and node connections for each Fog Node and WD for each time slot. It is noteworthy that in our model, the offloading is solely considered feasible from WDs to Fog Nodes, and the base station is solely regarded as the controlling entity.

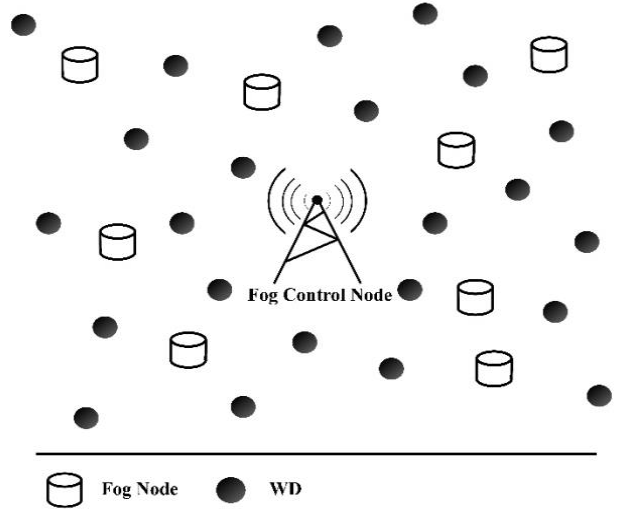

Fig. 1. Fog Network architecture

### A. Execution Model

The Fog Control Node schedules the CPU clock rate of each Fog Node based on various demands in each time slot. CPU clock rate of Fog Node j in each time slot t is defined as $f_j(t)$, $0 \leq f_j(t) \leq f^{max}, \forall j \in M$. And $f(t) \triangleq [f_1(t), \ldots, f_M(t)]$ denotes the CPU clock rate vector of all Fog Nodes.

The number of CPU cycles required to process a single bit task at the Fog Node j, $L_j$, depend on the types of tasks performed by WDs and can be obtained through offline measurements. The executed number of computation tasks at the Fog Node j in the time slot t, $\mu_j(t)$, can be formulated as:

$$\mu_j(t) = \tau f_j(t) L_j^{-1} \quad (1)$$

Furthermore, the power consumption for execution at the Fog Node j in time slot t is given by [18]:

$$P_j^c(t) = \kappa f_j^3(t) \quad (2)$$

The parameter K is contingent upon the deployed hardware and is quantifiable in practice.

We denote $\mu_i^j(t)$ as the number of computation tasks of the WD i that are executed at the Fog Node j in time slot t by a particular scheduling procedure, such as first in, first out (FIFO). Our scheduling policy ensures full efficiency, which means that the sum of executed computation tasks of the WDs in Fog Node j is equal to the executed number of computation tasks at Fog Node j, in each time slot [19]:

$$\sum_{i \in N} \mu_i^j(t) = \mu_j(t), \forall j \in M \quad (3)$$

### B. Computation Offloading and Wireless Transmission Model

We assume that each WD can transmit its computation tasks to any Fog Node with no restrictions. We also assume that each computation task on WDs are independent and can be run separately from other computation processes [20]–[23]. Thus, WD's computation tasks can be broke down into multiple chunks and each chunk can be offloaded to each Fog Node. We define the matrix element $\alpha_i^j(t)$ as the chunk of amount of WD i's computation task offloaded to Fog Node j in time slot t and is defined within [0, 1]. Define $\alpha(t) \triangleq [\alpha_1^1(t), ..., \alpha_N^1(t); ...; \alpha_1^M(t), ..., \alpha_N^M(t)]$ as the computation offloading indicator matrix. We assume that wireless devices use a space-division multiple access (SDMA) scheme to connect through wireless channel and each mobile Fog Node is equipped with R antennas. This means that each Fog Node can get tasks from at most R WDs simultaneously. These considerations define a set of constraints for the Fog Control Node scheduling algorithm, noted as:

$$0 \leq \alpha_i^j(t) \leq 1, \forall j \in M, i \in N \quad (4a)$$

$$\sum_{j \in M} \alpha_i^j(t) \leq 1, \forall i \in N \quad (4b)$$

$$\sum_{i \in N} \beta_i^j(t) \leq R, \forall j \in M \quad (4c)$$

$$\beta_i^j(t) = \begin{cases} 0, & \alpha_i^j(t) = 0 \\ 1, & \alpha_i^j(t) > 0 \end{cases} \quad (4d)$$

Where $\beta_i^j(t)$ is a binary variable defined to set task limits, as each Fog Node can acquire the tasks from a maximum number of R WDs simultaneously.

In section IV.D, we will show that constraint (4b) and (4c) will be simplified.

WDs transmit their Computation tasks through wireless connections to the Fog Nodes for execution. Assuming i.i.d. frequency-flat block fading wireless channels between the WDs and the Fog Nodes, we define the small-scale fading channel power gain from the WD i to the Fog Node j in the time slot t as $\sigma_i^j(t)$, that follows a Rayleigh distribution and is assumed to be constant in each time slot t. The channel power gain from the WD i to the Fog Node j can be represented by $G_i^j(t) = \sigma_i^j(t) g0 (d0/d_i^j(t))^\theta$, where $g0$ is the path-loss constant, $d0$ is the reference distance, $\theta$ is the path-loss exponent, and $d_i^j(t)$ is the distance between the mobile Fog Node j and WD i in time slot t.

According to SDMA utilization and Shannon's capacity formula [24], we derive that the amount of computation tasks that can be offloaded from WD i to Fog Node j in time slot t is given by:

$$C_i^j(t) = \omega\tau \log_2\left(1 + (P_i^{j,tr}(t)G_i^j(t))/(\omega N_0)\right), \forall j \in M, i \in N \quad (5)$$

where $P_i^{j,tr}(t)$ is the transmit power from WD i to Fog Node j with a maximum value $P_i^{j,max}$, that is scheduled by Fog Control Node for wireless transmission. $\omega$ is the available bandwidth at each Fog Node for each WD. $P^{tr}(t) \triangleq [P_1^{1,tr}(t), ..., P_N^{1,tr}(t); ...; P_1^{M,tr}(t), ..., P_N^{M,tr}(t)]$ is defined as a matrix consisted of all the transmission power values.

The number of computation tasks offloaded by WD i through wireless link to Fog Nodes in time slot t is formulated as:

$$\pi_i(t) = \sum_{j \in M} \alpha_i^j(t) C_i^j(t), \forall i \in N \quad (6)$$

*C. Queueing Model*

Let $A_i(t)$ (bits) denote the arriving computation tasks of WD i at the beginning of the time slot t, which can be processed at the next time slot (t+1). Without loss of generality, we assume the $A_i(t)$s in different time slots are independent and identically distributed (i.i.d.) within $[0, A_{i,max}]$. Each WD utilizes a buffer of a specified size to guarantee the availability of service for computation tasks that have not been offloaded. Consequently, it is capable of supporting only a portion of the received data based on the availability of its buffer. $a_i(t)$, i.e., $0 \leq a_i(t) \leq A_i(t) \leq A_{i,max}$. $a(t) \triangleq [a_1(t), ..., a_N(t)]$ is defined as a vector consisted of all the admitted data.

The arrived computations that have not yet been assigned to Fog Nodes will be queued in the task buffer at each WD with sufficient capacity [21], [22], [25], [26]. The backlog of the data queue at WD i at the beginning of the time slot t is denoted as $S_i(t)$, where $S_i(t)$ changes at each time slot according to the following routine:

$$S_i(t+1) = \left[S_i(t) - \sum_{j \in M} \alpha_i^j(t) C_i^j(t)\right]^+ + a_i(t), \forall i \in N \quad (7)$$

Where $[x]^+ = \max(x, 0)$ is a guarantee that the WD will not offload more data than the remaining data in its buffer when $S_i(t) < \sum_{j \in M} \alpha_i^j(t) C_i^j(t)$ occurs.

A task buffer is assumed to be available in each mobile Fog Node for each WD to store computations that have been offloaded but not yet executed. The task buffer is assumed to have sufficiently large capacity. $Q_i^j(t)$ is the queue length of Fog Node j's computation tasks where WD i has offloaded until the beginning of the time slot t. It changes as follows:

$$Q_i^j(t+1) = \left[Q_i^j(t) - \mu_i^j(t)\right]^+ + \min\{\alpha_i^j(t)C_i^j(t), \alpha_i^j(t)S_i(t)\}, \forall j \in M, i \in N \quad (8a)$$

The second term located on the right-hand side of (8a) indicates that the WD is unable to offload more data than what it has already stored. When $\alpha_i^j(t)C_i^j(t) > \alpha_i^j(t)S_i(t)$, the excessive offloading rate is used for transmitting dummy data without processing task overload [11].

Accordingly, due to our utilized fully efficient scheduling policy [19] that validates equation (3), data queue dynamics at mobile Fog Node j are defined as:

$$Q_j(t) = \sum_{i \in N} Q_i^j(t) \quad (8b)$$

$$Q_j(t+1) = \left[Q_j(t) - \mu_j(t)\right]^+ + \sum_{i \in N} \min\{\alpha_i^j(t)C_i^j(t), \alpha_i^j(t)S_i(t)\}, \forall j \in M \quad (8c)$$

III. PERFORMANCE METRICS AND PROBLEM FORMULATION

This section presents the performance metrics used to evaluate the system, which include the ratio of long-term utility to long-term total power and the time average mean of queue backlogs. The problem of utility-power efficiency maximization with task queues stability is then formulated, considering the computation offloading management, arrival data admission control, and resource scheduling.

Let $\overline{E_N}$ denote the time-average expected power consumption of WDs, i.e., $\overline{E_N} = \lim_{t\to\infty} \frac{1}{t}\sum_{k=0}^{t-1} \mathbb{E}\{\sum_{i\in N}\sum_{j\in M} P_i^{j,tr}(k)\}$, where $\mathbb{E}\{.\}$ is the expectation operator. Similarly, let $\overline{E_M}$ denote the time-average expected power consumption of all Fog Nodes, i.e., $\overline{E_M} = \lim_{t\to\infty} \frac{1}{t}\sum_{k=0}^{t-1} \mathbb{E}\{\sum_{j\in M} P_j^c(k)\}$, and let $\overline{a_\iota}$ denote the throughput of the i-th WD, i.e., $\overline{a_\iota} = \lim_{t\to\infty} \frac{1}{t}\mathbb{E}\{\sum_{k=0}^{t-1} a_i(k)\}$, for all $i \in N$. These quantities are computed over the network states, random task arrivals, and scheduling policies. Note that all expectations are taken with respect to the random network state information, task arrivals and the policies, throughout this paper. The utility function of WD i, denoted by $U_i(\overline{a_\iota})$, is a differentiable, increasing, and concave function of the throughput $\overline{a_\iota}$ achieved by the i-th WD. It represents the satisfaction level of WD i for having a throughput of $\overline{a_\iota}$. Accordingly, the utility-power efficiency performance metric is defined as $\eta = \frac{\sum_{i\in N} U_i(\overline{a_\iota})}{\overline{E_M}+\overline{E_N}+Co}$, where $Co$ is a positive constant that represents the power consumption of the Fog Control Node. This metric reflects how efficiently the system can deliver utility with respect to power consumption. The power consumption of the Fog Control Node is assumed to be negligible compared to the total power consumption of the system.

According to Little's Law [27], the average execution delay experienced by WDs is proportional to the average number of their tasks waiting in the fog network, which is the sum of the task buffer queue lengths at the WDs and Fog Nodes sides. Therefore, the time average mean of queue backlog of the task buffers for WDs and Fog Nodes is used as a measure of the execution delay, which can be expressed as:

$$D = \lim_{t\to\infty}\frac{1}{t}\sum_{k=0}^{t-1}\mathbb{E}\left\{\frac{\sum_{j\in M}Q_j(k)}{|M|} + \frac{\sum_{i\in N}S_i(k)}{|N|}\right\} \quad (9)$$

The utility-power efficiency maximization problem is formulated as:

$$P1: \max_\chi \eta = \frac{\sum_{i\in N} U_i(\overline{a_\iota}(\chi))}{\overline{E_M}(\chi)+\overline{E_N}(\chi)+Co}$$

$$st \quad (4a)(4b)(4c)$$

$$0 \leq f_j(t) \leq f^{max}, \forall j \in M \quad (10a)$$

$$0 \leq P_i^{j,tr}(t) \leq P_i^{j,max}, \forall i \in N, \forall j \in M \quad (10b)$$

$$0 \leq a_i(t) \leq A_i(t) \leq A_{i,max}, \forall i \in N \quad (10c)$$

$$\lim_{t\to\infty}\frac{1}{t}\sum_{k=0}^{t-1}\mathbb{E}\{Q_j(k)\} < \infty, \forall j \in M \quad (10d)$$

$$\lim_{t\to\infty}\frac{1}{t}\sum_{k=0}^{t-1}\mathbb{E}\{S_i(k)\} < \infty, \forall i \in N \quad (10e)$$

where (10a), (10b) and (10c) are the CPU clock rate constraint, transmit power constraint and the arrival data constraint, respectively and (4a)-(4c) are the computation offloading indicator constraint. (10d) and (10e) ensure the stability of all the data queues [28], which ensure that all the arrived computation tasks can be executed within finite time. Also, $\chi(t) = (f(t), P^{tr}(t), \alpha(t), a(t))$ is the policy for time slot t, which are the control variables in our system. And let $\chi = (\chi(0), \chi(1), ...)$ denote a policy.

The problem (P1) is challenging due to the fractional form of objective function and usual Lyapunov optimization techniques cannot be adopted in a straightforward manner. To this end, denote $\chi_{P1}^*$ and $\eta_{P1}^*$ as the optimal policy for (P1) and corresponding optimal utility-power efficiency, i.e., $\eta_{P1}^* = \frac{\sum_{i\in N} U_i(\overline{a_\iota}(\chi_{P1}^*))}{\overline{E_M}(\chi_{P1}^*)+\overline{E_N}(\chi_{P1}^*)+Co}$, we can eliminate the fractional form of utility-power efficiency function by the following theorem.

***Theorem 1:*** The optimal policy $\chi_{P1}^*$ achieves the optimal utility-power efficiency $\eta_{P1}^*$ if and only if

$$\max_\chi \sum_{i\in N} U_i(\overline{a_\iota}(\chi_{P1})) - \eta_{P1}^*[\overline{E_M}(\chi_{P1}) + \overline{E_N}(\chi_{P1}) + Co] = \sum_{i\in N} U_i(\overline{a_\iota}(\chi_{P1}^*)) - \eta_{P1}^*[\overline{E_M}(\chi_{P1}^*) + \overline{E_N}(\chi_{P1}^*) + Co] = 0 \quad (11)$$

Proof. The proof is the same as theorem 1 of Zhang et al. [29] technical report. □

Thus, (P1) can be equivalently reformulated as

$$P2: \max_\chi \sum_{i\in N} U_i(\overline{a_\iota}(\chi_{P1})) - \eta_{P1}^*[\overline{E_M}(\chi_{P1}) + \overline{E_N}(\chi_{P1}) + Co]$$

$$st \quad (4a) - (4c), (10a) - (10e)$$

Theorem 1 shows that the optimal value of (P2) is zero. We assume $\eta_{P1}^*$ to be known, but we will relax this assumption later in the development of our algorithm.

In order to process the objective function as a slot-by-slot approach and utilize the Lyapunov drift-plus-penalty framework, we use the following lemma to smooth the utility term, which is a concave function of time average, and hence formulate a relaxed problem.

***Lemma 1:*** P2 can be equivalently reformulated as

$$P3: \max_{\chi,\gamma} \sum_{i\in N} \overline{U_\iota(\gamma_\iota)} - \eta_{P1}^*[\overline{E_M}(\chi_{P1}) + \overline{E_N}(\chi_{P1}) + Co]$$

$$st \quad (4a) - (4c), (10a) - (10e)$$

$$0 \leq \gamma_i(t) \leq A_{i,max}, \forall t, \forall i \in N \quad (12a)$$

$$\overline{\gamma_\iota} \leq \overline{a_\iota}, \forall i \in N \quad (12b)$$

where $\gamma(t) = (\gamma_1(t), ..., \gamma_N(t))$ is the defined auxiliary variables.

Proof. See appendix A. □

By maximizing the time-average of functions rather than the function of time-averages in (P3), we facilitate the use of the Lyapunov drift-plus-penalty framework.

In general, (P3) is a stochastic optimization problem with various stochastic information to be handled, such as the CPU clock rate scheduling for each mobile Fog Node, the computation offloading, the data admission, and the transmission power allocation for each WD, in each time slot. It requires a prior knowledge of network information, such as task traffic, arrival data and available computation resources. This information is difficult to predict as our model considers the node mobility. To address this challenge, we employ the Lyapunov optimization technique, which can obtain the optimal solution with a practical complexity, and we can solve the challenging stochastic optimization problem (P3) by solving a deterministic problem for each time slot in that time frame. In the next section, we will show that our algorithm can achieve asymptotic optimality and reveal the trade-off between utility-power efficiency and average mean of queue backlogs in fog network.

## IV. ONLINE AND UTILITY-POWER EFFICIENT TASK SCHEDULING ALGORITHM

In this section, we present an online and utility-power efficient task scheduling algorithm to address P3 based on the Lyapunov optimization technique [28]. To this end, we first define the virtual queue $Z_i(t)$ for each $i \in N$, which is used to ensure the satisfaction of the constraint (12b). The virtual queue evolves according to the following update rule:

$$Z_i(t+1) = [Z_i(t) + \gamma_i(t) - a_i(t)]^+ \quad (13)$$

We define $Z(t) \triangleq [Z_1(t), \ldots, Z_N(t)]$ as the vector of virtual queues and the quadratic Lyapunov function as:

$$L(\Theta(t)) \triangleq \tfrac{1}{2}\sum_{j \in M} Q_j^2(t) + \tfrac{1}{2}\sum_{i \in N} S_i^2(t) + \tfrac{1}{2}\sum_{i \in N} Z_i^2(t) \quad (14)$$

Where $\Theta(t)$ is a vector that concatenates all actual and virtual queues, $\Theta(t) = [Q^T(t), S^T(t), Z^T(t)]$. We also introduce the conditional Lyapunov drift function $\Delta(\Theta(t))$ as:

$$\Delta(\Theta(t)) \triangleq \mathbb{E}\{L(\Theta(t+1)) - L(\Theta(t))|\Theta(t)\} \quad (15)$$

Which measures the change of the Lyapunov function over time slots. Therefore, the one-slot conditional Lyapunov drift-plus-penalty function is given by:

$$\Delta_v(\Theta(t)) = \Delta(\Theta(t)) - V\{\sum_{i \in N} \mathbb{E}\{U_i(\gamma_i(t))|\Theta(t)\} - \eta_{p1}^*[\sum_{j \in M} \mathbb{E}\{P_j^c(t)|\Theta(t)\} + \sum_{j \in M}\sum_{i \in N} \mathbb{E}\{P_i^{j,tr}(t)|\Theta(t)\} + Co]\} \quad (16)$$

where V is a non-negative control parameter that balances the trade-off between queue backlog stability and utility-power efficiency.

We note that the optimal utility-power efficiency $\eta_{p1}^*$ is unknown, so we use the running averages of the power consumption and the auxiliary variable $\gamma(t)$ defined as (17) as an approximation, and update the conditional Lyapunov drift-plus-penalty function every time slot using $\eta(t)$. We prove the convergence of $\eta(t)$ to $\eta_{p1}^*$ in Appendix B.

$$\eta(t) = \frac{\sum_{i \in N} U_i\left(\tfrac{1}{t}\sum_{\tau=0}^{t-1}\gamma_i(\tau)\right)}{\left(\tfrac{1}{t}\sum_{\tau=0}^{t-1} E_M(\tau) + E_N(\tau)\right) + Co} \quad (17)$$

The following lemma establishes an upper bound for the drift-plus-penalty function.

***Lemma 2:*** The one-slot conditional Lyapunov drift-plus-penalty function is upper bounded for any feasible policy χ(t) that can be applied in any time slot t, i.e.,

$$\Delta_v(\Theta(t)) \leq \vartheta - V\{\sum_{i \in N} \mathbb{E}\{U_i(\gamma_i(t))|\Theta(t)\} - \mathbb{E}\{\eta(t) * [\sum_{j \in M} P_j^c(t) + \sum_{j \in M}\sum_{i \in N} P_i^{j,tr}(t) + Co]|\Theta(t)\}\} + \mathbb{E}\{\sum_{j \in M} Q_j(t)[\sum_{i \in N}\alpha_i^j(t)C_i^j(t) - \mu_j(t)]|\Theta(t)\} + \mathbb{E}\{\sum_{i \in N} Z_i(t)[\gamma_i(t) - a_i(t)]|\Theta(t)\} + \mathbb{E}\{\sum_{i \in N} S_i(t)[a_i(t) - \sum_{j \in M}\alpha_i^j(t)C_i^j(t)]|\Theta(t)\} \quad (18)$$

where $\vartheta$ is a positive constant.

Proof: Please refer to Appendix C. □

The main idea of our online and utility-power efficient task scheduling algorithm is to minimize the upper bound of $\Delta_v(\Theta(t))$ in the right-side of (18) at each time slot. By doing so, the backlog of task queues can be kept short, while the utility-power efficiency of the fog network can be maximized. In each time slot, we take the conditional expectation of the right-side of (18) with respect to the random network state information and the policy χ(t). Our online task scheduling algorithm observes the current queues Θ(t), network system information, and updates $\eta(t)$ according to (17). Thus, it does not need any information on the distribution or future values of network system states. Therefore, we transform the utility-power efficiency maximization problem P3 into the queue backlog stability and utility-power efficiency balanced optimization problem P4 in each time slot.

$$P4: \min_{\chi(t),\gamma(t)} -V\left\{\begin{array}{l}\sum_{i \in N} U_i(\gamma_i(t)) - \\ \eta(t)\left[\sum_{j \in M} P_j^c(t) + \sum_{j \in M}\sum_{i \in N} P_i^{j,tr}(t)\right]\end{array}\right\} + \sum_{j \in M} Q_j(t)\left[\sum_{i \in N}\alpha_i^j(t)C_i^j(t) - \mu_j(t)\right] + \sum_{i \in N} Z_i(t)[\gamma_i(t) - a_i(t)] + \sum_{i \in N} S_i(t)\left[a_i(t) - \sum_{j \in M}\alpha_i^j(t)C_i^j(t)\right]$$

$$st \quad (4a)-(4c), (10a)-(10c), (12a)$$

P4 can be decomposed into four sub-problems by eliminating the coupling terms: A) The problem of determining the auxiliary variable $\gamma(t)$, B) The problem of deciding on the data admission in WDs $a(t)$ that affects the amount of data to be processed and transmitted, C) The problem of setting the CPU clock rate of each Fog Node $f(t)$ that controls the processing capability and the power consumption of the Fog Nodes, D) The problem of jointly optimizing the transmission power and the offloading factors for WDs $P^{tr}(t)$ and $\alpha(t)$, that influence the communication energy and the delay of the WDs. The analysis of these sub-problems is presented in the following sections.

### A. Auxiliary Variable $\gamma(t)$ Determination:

$$sp1: \min_{\gamma(t)} -V\sum_{i \in N} U_i(\gamma_i(t)) + \sum_{i \in N} Z_i(t)\gamma_i(t)$$

$$st \ (12a)$$

The problem can be solved separately for each WD and is convex. Therefore, the optimal solution can be expressed as:

$$\gamma_i(t) = \left[(U_i')^{-1}\left(\frac{Z_i(t)}{V}\right)\right]_0^{A_{i,max}}, \forall i \in N \quad (19)$$

The virtual queue length $Z_i(t)$ decreases when the system has enough capacity and satisfies (12b). According to equation (19), $\gamma_i(t)$ increases to accept more computation tasks by increasing $Z_i(t)$. This is due to the fact that $(U_i')^{-1}(.)$ is a decreasing function.

B. *Data Admission Decision In WDs $a(t)$:*

$$sp2: \min_{a(t)} \sum_{i \in N} a_i(t)[S_i(t) - Z_i(t)]$$

$$st\ (10c)$$

By comparing the queue backlogs, each $a_i(t)$ is independently calculated as follows:

$$a_i(t) = \begin{cases} A_i(t), & S_i(t) - Z_i(t) < 0 \\ 0, & S_i(t) - Z_i(t) \geq 0 \end{cases}, \forall i \in N \quad (20)$$

C. *CPU Clock Rate of Fog Nodes $f(t)$ Scheduling:*

$$sp3: \min_{f(t)} V\eta(t) \sum_{j \in M} \kappa f_j^3(t) - \sum_{j \in M} Q_j(t)(\tau f_j(t) L_j^{-1})$$

$$st\ (10a)$$

The sp3 problem is a convex optimization problem. The optimal CPU clock rate scheduling for each Fog Node $j \in M$, which minimizes the power consumption and delay, can be obtained by solving the following equation:

$$f_j(t) = \left[\sqrt{\frac{Q_j(t)\tau}{3\kappa V\eta(t)L_j}}\right]_{f^{min}}^{f^{max}}, \forall j \in M \quad (21)$$

D. *Joint Power Transmission Allocation of WDs $P^{tr}(t)$ and Computation Offloading $\alpha(t)$*

In this subproblem, we have two sets of optimization variables: the computation offloading indicator and the transmission power allocation of WDs.

$$sp4: \min_{P^{tr}(t),\alpha(t)} \left\{ V\eta(t) \sum_{j \in M} \sum_{i \in N} P_i^{j,tr}(t) \right. $$
$$+ \sum_{j \in M} Q_j(t) \sum_{i \in N} \alpha_i^j(t) \omega\tau \log_2(1 $$
$$+ (P_i^{j,tr}(t) G_i^j(t))/(\omega N_0)) $$
$$- \sum_{i \in N} S_i(t) \sum_{j \in M} \alpha_i^j(t) \omega\tau \log_2(1 $$
$$\left. + (P_i^{j,tr}(t) G_i^j(t))/(\omega N_0)) \right\}$$

$$st\ (4a) - (4c), (10b)$$

Given that the feasible region of this subproblem is a Cartesian product of the feasible regions of $P^{tr}(t)$ and $\alpha(t)$, and is known to be convex, the Gauss-Seidel method can be employed as a time-efficient approach to solving this optimization problem [30], [31]. By utilizing this method, the power transmission of wireless devices (WDs) and the computation offloading indicators can be optimized alternately in an iterative manner. At each iteration, a fixed and updated computation offloading indicator and allocated power transmission are used. The application of this iterative process ensures convergence towards the global optimal solution of problem sp4.

With a fixed computation offloading indicator $\alpha(t)$, the power transmission of each WD $i \in N$ to Fog Node $j \in M$ is given by:

$$P_i^{j,tr}(t) = min\left(P_i^{j,max}, \left[\left(\frac{(S_i(t) - Q_j(t))\alpha_i^j(t)\omega\tau}{V\eta(t)\ln(2)}\right) - \left(\frac{\omega N_0}{G_i^j(t)}\right)\right]^+\right) \quad (22)$$

For fixed transmission power allocation, the following computation offloading problem can be formulated as:

$$sp\_\alpha: \min_{\alpha(t)} \left\{ \sum_{j \in M} Q_j(t) \sum_{i \in N} \alpha_i^j(t) C_i^j(t) \right.$$
$$\left. - \sum_{i \in N} S_i(t) \sum_{j \in M} \alpha_i^j(t) C_i^j(t) \right\}$$

$$st\ (4a) - (4c)$$

***Theorem 2:*** Optimal computation offloading variables $\alpha(t)$ in $sp\_\alpha$ are binary, i.e., they take values of either 0 or 1.

Proof : Please refer to Appendix D. □

By Theorem 2, we can conclude that the computation offloading variables are binary. Therefore, we can simplify the computation offloading problem as follows:

$$sp\_\alpha: \min_{\alpha(t)} \left\{ \sum_{j \in M} Q_j(t) \sum_{i \in N} \alpha_i^j(t) C_i^j(t) \right.$$
$$\left. - \sum_{i \in N} S_i(t) \sum_{j \in M} \alpha_i^j(t) C_i^j(t) \right\}$$

$$st\ (4a)$$

$$\sum_{j \in M} \alpha_i^j(t) \leq 1, \forall i \in N \quad (23a)$$

$$\sum_{i \in N} \alpha_i^j(t) \leq R, \forall j \in M \quad (23b)$$

Thus, we can apply convex optimization techniques to find the optimal task offloading decision. Note that, by Theorem 2, the allocated bandwidth $\omega$ will be fully utilized by each WD.

V. LONG-TERM THEORETICAL PERFORMANCE ANALYSIS

In this section, we present a theoretical performance analysis of the online and utility-power efficient task scheduling algorithm, focusing on the metrics of utility-power efficiency and average mean of queue backlogs. This analysis establishes both the upper bound for the average mean of queue backlogs and the lower bound for the utility-power efficiency, thereby revealing the trade-off relationship between these metrics.

To conduct the performance analysis, we assume an i.i.d. system randomness with a stationary distribution. Additionally, we assume the existence of values $\varepsilon > 0$, a policy characterized by a fixed vector $\gamma'$, a stationary randomized policy $\chi'(t)$, and an utility-power efficiency $\eta'$ achieved by the combination of $(\chi'(t), \gamma')$. It is important to note that these values and policies are not required to be optimal; they simply need to satisfy the following Slater-type conditions:

$$\mathbb{E}\{\sum_{i \in N} \alpha_i^j(\chi'(t)) C_i^j(\chi'(t)) - \mu_j(\chi'(t))\} \leq -\varepsilon, \forall j \in M \quad (24)$$

$$\mathbb{E}\{a_i(\chi'(t)) - \sum_{j \in M} \alpha_i^j(\chi'(t)) C_i^j(\chi'(t))\} \leq -\varepsilon, \forall i \in N \quad (25)$$

$$\gamma'_i - \mathbb{E}\{a_i(\chi'(t))\} \leq 0, \forall i \in N \quad (26)$$

$$\sum_{i \in N} U_i(\gamma'_i) - \eta'[\sum_{j \in M} \mathbb{E}\{P_j^c(\chi'(t))\} + \sum_{j \in M} \sum_{i \in N} \mathbb{E}\{P_i^{j,tr}(\chi'(t))\} + Co] \geq 0 \quad (27)$$

Assumptions (24)-(26) guarantee the strong stability of the queues $Q_j(t)$, $S_i(t)$ and $Z_i(t)$, which are commonly assumed in network stability analysis literature [28]. These assumptions ensure the existence of a stationary and randomized task scheduling algorithm for a homogeneous fog network.

Based on the aforementioned Slater-type conditions, the performance bounds of utility-power efficiency and average mean of queue backlogs for our online task scheduling algorithm are derived in the following theorem.

***Theorem 3:*** For homogeneous fog network defined in section II, the online task scheduling algorithm achieves the following properties:

*1) The utility-power efficiency is bounded by:*
$$\eta \geq \eta_{P2}^* - \frac{\vartheta}{VCo} \quad (28)$$

*2) The performance bound of the average mean of queue backlogs is given by:*
$$D \leq \frac{\vartheta}{\varepsilon} + V \frac{\eta e_{max}}{\varepsilon} \quad (29)$$

Proof : Please refer to Appendix E. □

Theorem 3 demonstrates that under our online task scheduling algorithm, the lower bound of utility-power efficiency increases inversely proportional to V, while the upper bound of the average mean of queue backlogs increases linearly with V. Therefore, there exists an $[O(1/V), O(V)]$ trade-off between these two metrics.

## VI. EVALUATION

To demonstrate the feasibility, convergence, and stability of our online and utility-power efficient task scheduling algorithm over time, we conducted simulations using the following fog network configurations. The simulations were performed once over 10,000 constant time slots.

The fog network consisted of randomly deployed $|M| = 8$ mobile Fog Nodes and $|N| = 40$ low mobility WDs, covering an area of $150 * 150\ m^2$. The mobility of nodes and devices was modeled using random waypoint node mobility [32]. The parameters used in the simulations were as follows: $\mathbb{E}\{\sigma_i^j(t)\} = 1$, $g0 = -40dB$, $\theta = 5$, $d0 = 1m$, $N0 = -174\ dBm/Hz$, $\omega = 10\ MHz$, $\kappa = 10^{-27}$, $Lj = 500 cycles/bit$, $P_i^{j,max} = 200mW$, $f^{max} = 2GHz$, $Co = 64$, R=3. The arrival tasks $A_i(t)$ were uniformly distributed within the range $[0, 4kbit]$, and $\tau$ was set to $1\ ms$.

Figs. 2-4 were generated using the aforementioned settings. In Fig. 2, the convergence of utility-power efficiency $\eta(t)$ over time is depicted. This utility-power efficiency was used to solve the updated problem, which changed with each time slot based on the new $\eta(t)$ (18). Fig. 3 illustrates the mean of real queue backlogs ($Q_j$ and $S_i$) over time slots. It can be observed that the mean of real queue backlogs converge to a finite value, indicating the mean rate stability of real queues (10d) and (10e).

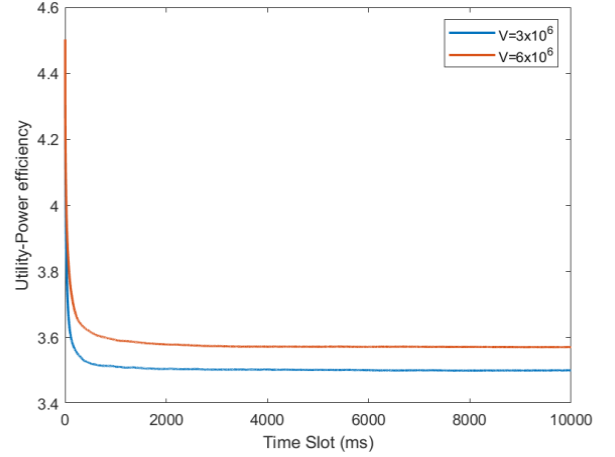

Fig. 2. Convergence of utility-power efficiency under two different V values.

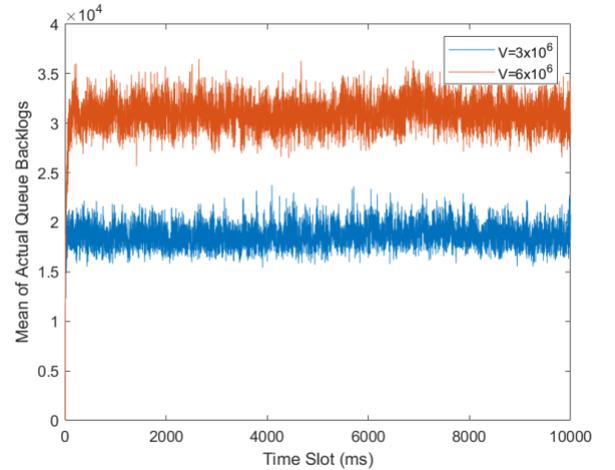

Fig. 3. Convergence of mean of actual queue backlogs under two different V values.

To assess the performance of our algorithm under different values of the trade-off parameter V, we conducted simulations using the same fog network settings mentioned earlier. The simulation results were averaged over 10,000 time slots. Through these simulations, we showcased the achieved trade-off between utility-power efficiency and average mean of queue backlogs.

As depicted in Fig. 4, the utility-power efficiency is directly proportional to V and reaches its optimal value when V is sufficiently large. However, this increase in utility-power efficiency is accompanied by a corresponding increase in delay, as predicted by equations (28) and (29) in Theorem 3. Furthermore, as indicated by equation (29), the average mean of queue backlogs grows linearly with V. These observations confirm the theoretical results of Theorem 3 and demonstrate the trade-off $[O(1/V), O(V)]$ between utility-power efficiency and average mean of queue backlogs achieved by our proposed algorithm. In other words, a higher value of V improves utility-power efficiency while introducing higher average mean of queue backlogs, i.e., delay.

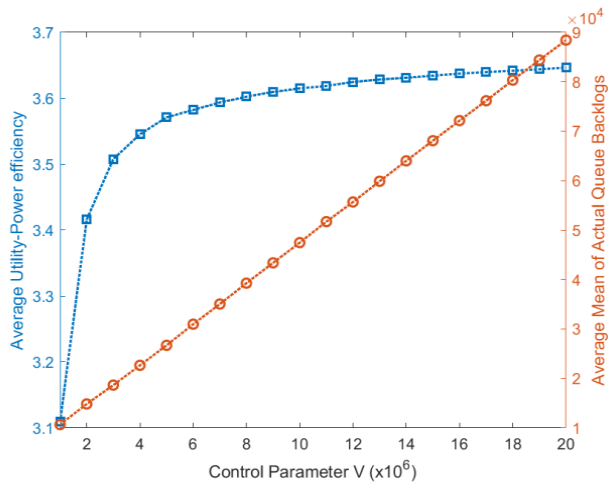

Fig. 4. Utility-power efficiency and average mean of actual queue backlogs performance

Fig. 4 illustrates that the optimal trade-off region for V lies between $10^6$ and $7 * 10^6$. Beyond this region, the efficiency does not exhibit significant improvement, but the queue backlogs increase linearly.

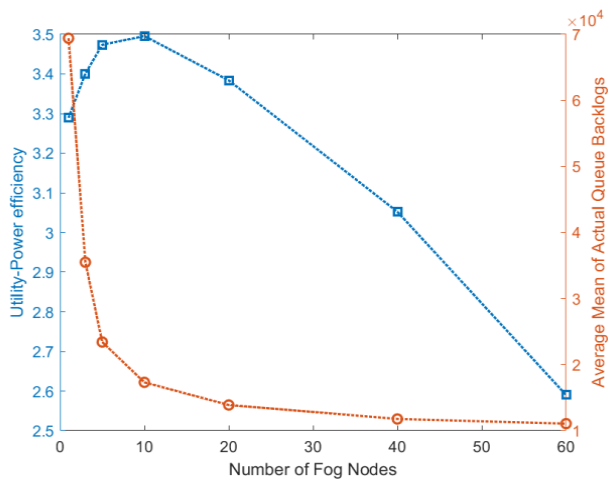

Fig. 5. Proposed algorithm performance with Fog Nodes incremented, for $|N| = 40$ and $V = 3 * 10^6$.

Fig. 5 presents the performance evaluation of our algorithm under varying numbers of Fog Nodes, using the same settings as described earlier and fixing V at $3 * 10^6$. As the number of Fog Nodes increases, the utility initially rises until reaching a threshold where the maximum number of tasks can be efficiently offloaded. Beyond this threshold, there is a marginal reduction in the average mean of queue backlogs accompanied by increased power consumption. The utility-power efficiency exhibits a concave pattern, with an initial increase attributed to the Fog Nodes accommodating more WDs. However, beyond the peak, the utility-power efficiency declines as the Fog Nodes struggle to handle additional WDs, resulting in a lower rate of utility growth compared to power consumption. This concave pattern reflects the saturation of the optimal capacity of the Fog Nodes.

Fig. 6 illustrates the performance of our algorithm as a function of the number of WDs. As the number of WDs increases, the algorithm selects the most optimal WDs to offload their tasks to the existing set of fog nodes, thereby reducing the network's power consumption and increasing the admitted data in proportion. Consequently, the utility-power efficiency exhibits a linear increase. The average mean of queue backlogs also increases up to a certain point, which depends on several factors: 1) the maximum number of WDs capable of offloading their tasks, considering the limited number of antennas available for each fog node, 2) the maximum size of the tasks arriving at each WD, and 3) the chosen control parameter V, which balances the average mean of queue backlogs and utility-power efficiency.

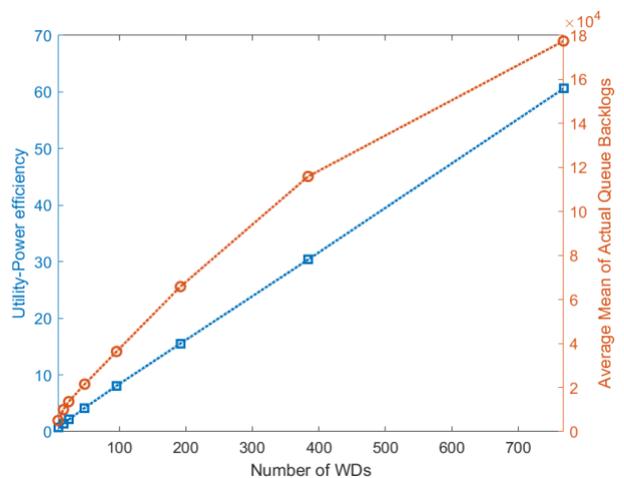

Fig. 6. Proposed algorithm performance with WDs incremented, for $|M| = 8$ and $V = 3 * 10^6$

## VII. CONCLUSIONS

In this study, we addressed the problem of online and utility-power efficient task scheduling within a fog network comprising homogeneous mobile Fog Nodes and low-mobility wireless devices, all controlled by a Fog Control Node. The objective was to offload computation tasks from the wireless devices (WDs) to nearby Fog Nodes with available computation resources, in an optimal and efficient manner.

To tackle this problem, we formulated it as a stochastic optimization problem and subsequently modified the general Lyapunov optimization technique. This modification allowed us to derive a short-term problem that incorporated the unknown optimal utility-power efficiency. Based on this framework, we proposed a novel and online utility-power efficient task scheduling algorithm. The algorithm encompassed computation offloading management, arrival data admission control, and resource scheduling.

The performance of our algorithm was evaluated through theoretical analysis and simulations. These evaluations explicitly characterized the trade-offs between utility-power efficiency and average mean of queue backlogs. The results demonstrated that our proposed algorithm achieved both fair throughput among the wireless devices and reduction in power consumption for both the Fog Nodes and the wireless devices.

In conclusion, our study contributes to the field of fog network task scheduling by providing a novel algorithm that addresses the challenges of utility-power efficiency in an online setting. The theoretical analysis and simulation results validate the effectiveness of our approach in achieving the desired trade-offs and optimizing the performance of the system.

## VIII. APPENDIX

### A. Proof of Lemma 1

Indicate $\sum_{i \in N} \overline{U_i(\gamma'_{i,P3})}$ and $\overline{a'_{i,P3}}$ as the maximum utility for P3 and corresponding throughput, according to (12b) and exerted monotonically increasing utility function, $\sum_{i \in N} U_i(\overline{a'_{i,P3}}) \geq \sum_{i \in N} U_i(\overline{\gamma'_{i,P3}})$ holds true. Also, according to Jensen's inequality we have $\sum_{i \in N} \overline{U_i(\gamma'_{i,P3})} \leq \sum_{i \in N} U_i(\overline{\gamma'_{i,P3}})$. Indicating $a^*_{i,P2}(t)$ as the optimal admitted data of P2 in t'th time slot, under (12a) and (12b) constraints, and applying the strategy $\overline{a^*_{i,P2}(t)} = \gamma_i(t)$ for each time slot and WD, attains $\sum_{i \in N} \overline{U_i(\gamma_i(t))} = \sum_{i \in N} U_i(\overline{a^*_{i,P2}(t)})$. Consequently we have $\sum_{i \in N} U_i(\overline{a'_{i,P3}}) \geq \sum_{i \in N} U_i(\overline{a^*_{i,P2}(t)})$. Thus P3 is equivalent to P2.

### B. Proof of Convergence of $\eta(t)$ To $\eta^*_{p1}$

Whenever $\eta(t)$ get bigger than $\eta^*_{p1}$ then modified Lyapunov drift-plus-penalty function (30) aims to maximize $\sum_{i \in N} U_i(\bar{a}_i) - \eta(t)[\overline{E_M} + \overline{E_N} + Co]$ in a long-term meaning. Due to (11), we have $\sum_{i \in N} U_i(\bar{a}_i) - \eta(t)[\overline{E_M} + \overline{E_N} + Co] < \sum_{i \in N} U_i(\bar{a}_i) - \eta^*_{p1}[\overline{E_M} + \overline{E_N} + Co] \leq 0$, thus the utility-power efficiency, $\frac{\sum_{i \in N} U_i(\bar{a}_i)}{\overline{E_M} + \overline{E_N} + Co}$ will be less than $\eta(t)$. Consequently, $\eta(t)$ reduces after t'th time slot. Under other conditions, if $\eta(t)$ become smaller than $\eta^*_{p1}$, due to the similar reason, $\eta(t)$ gets bigger after t'th time slot. Thus, $\eta(t)$ to come closer to $\eta^*_{p1}$ with each time slot.

$$\Delta_v(\Theta(t)) = \Delta(\Theta(t)) - V\{\sum_{i \in N} \mathbb{E}\{U_i(\gamma_i(t))|\Theta(t)\} - \mathbb{E}\{\eta(t) * [\sum_{j \in M} P^c_j(t) + \sum_{j \in M} \sum_{i \in N} P^{j,tr}_i(t) + Co] |\Theta(t)\}\} \quad (30)$$

### C. Proof of Lemma 2

Quadratic Lyapunov function definition will be exercised to calculate the upper bound of the development of one-slot conditional Lyapunov drift function. Also, the fact that for any positive a, b, c, (31) holds true, will be utilized.

$$([a-b]^+ + c)^2 \leq a^2 + c^2 + b^2 + 2a(c-b) \quad (31)$$

$$L(\Theta(t+1)) - L(\Theta(t)) = \frac{1}{2}\sum_{j \in M}[Q^2_j(t+1) - Q^2_j(t)] + \frac{1}{2}\sum_{i \in N}[S^2_i(t+1) - S^2_i(t)] + \frac{1}{2}\sum_{i \in N}[Z^2_i(t+1) - Z^2_i(t)] =$$
$$\frac{1}{2}\sum_{j \in M}\left[\left([Q_j(t) - \mu_j(t)]^+ + \sum_{i \in N} min\{\alpha^j_i(t) C^j_i(t), \alpha^j_i(t) S_i(t)\}\right)^2 - Q^2_j(t)\right] +$$
$$\frac{1}{2}\sum_{i \in N}\left[\left([S_i(t) - \sum_{j \in M} \alpha^j_i(t) C^j_i(t)]^+ + a_i(t)\right)^2 - S^2_i(t)\right] + \frac{1}{2}\sum_{i \in N}[([Z_i(t) + \gamma_i(t) - a_i(t)]^+)^2 - Z^2_i(t)] \leq$$

$$\frac{1}{2}\sum_{j \in M}\left[\left([Q_j(t) - \mu_j(t)]^+ + \sum_{i \in N}\left(\alpha^j_i(t) C^j_i(t)\right)\right)^2 - Q^2_j(t)\right] + \frac{1}{2}\sum_{i \in N}\left[\left([S_i(t) - \sum_{j \in M} \alpha^j_i(t) C^j_i(t)]^+ + a_i(t)\right)^2 - S^2_i(t)\right] + \frac{1}{2}\sum_{i \in N}[([Z_i(t) + \gamma_i(t) - a_i(t)]^+)^2 - Z^2_i(t)] \leq \frac{1}{2}\sum_{j \in M}\left[\left(\mu_j(t)\right)^2 + \left(\sum_{i \in N}\left(\alpha^j_i(t) C^j_i(t)\right)\right)^2\right] + \frac{1}{2}\sum_{i \in N}\left[\left(\sum_{j \in M} \alpha^j_i(t) C^j_i(t)\right)^2 + (a_i(t))^2\right] + \frac{1}{2}\sum_{i \in N}\left[(a_i(t) - \gamma_i(t))^2\right] + \sum_{j \in M} Q_j(t)\left[\sum_{i \in N} \alpha^j_i(t) C^j_i(t) - \mu_j(t)\right] + \sum_{i \in N} S_i(t)\left[a_i(t) - \sum_{j \in M} \alpha^j_i(t) C^j_i(t)\right] + \sum_{i \in N} Z_i(t)[\gamma_i(t) - a_i(t)] \quad (32)$$

Performing the conditional expectation of (32) and adding the penalty term to both sides, the Lemma 2 will be proven.

### D. Proof of Theorem 2

The solution of a minimization (maximization) problem with linear objective function and constraints is the boundary point of the feasible region. Therefore, since the relationship between objective function and computation offloading variables are linear to minimize the objective functions, they just adopt the boundary amounts. Since (4a) constraint hold true, they can be either 0 or 1.

### E. Proof of Theorem 3

To continue with the proof, we first introduce the following Lemma.

**Lemma 3:** under the convergence assumptions (33)-(34), we have (35)-(36).

$$\lim_{T \to \infty} \frac{1}{T}\sum_{t=0}^{T-1}[E_M(t) + E_N(t)] = e^{av} \quad (33)$$

$$\lim_{T \to \infty} \frac{1}{T}\sum_{t=0}^{T-1}\gamma_i(t) = \gamma^{av}_i \quad (34)$$

$$\lim_{T \to \infty} \frac{1}{T}\sum_{t=0}^{T-1}\mathbb{E}\{(E_M(t) + E_N(t) + Co)\eta(t)\} = \sum_{i \in N} U_i(\overline{\gamma_i}) \quad (35)$$

$$\lim_{T \to \infty} \frac{1}{T}\sum_{t=0}^{T-1}\mathbb{E}\{\eta(t)\} \leq \eta \quad (36)$$

Where $\eta$ is the solution to P1. The proof is the same as Zhang et al. [29] technical report.

We first prove the second part of Theorem 3. We denote the online and utility-power efficient task scheduling policy, as $(\chi^\xi(t), \gamma^\xi)$, which meets the Slater type condition and only depends on i.i.d. system state with stationary distribution. By exercising the Slater type condition to the right-side of (18) and arranging the terms, we have

$$\Delta(\Theta(t)) - V\mathbb{E}\left\{\begin{matrix}\sum_{i \in N} U_i(\gamma_i(t)) - \eta(t) * \\ [\sum_{j \in M} P^c_j(t) + \sum_{j \in M}\sum_{i \in N} P^{j,tr}_i(t) + Co]\end{matrix}\bigg|\Theta(t)\right\} \leq \vartheta - V\{\sum_{i \in N} U_i(\gamma^\xi_i) - \mathbb{E}\{\eta^\xi(t) * [\sum_{j \in M} P^c_j(\chi^\xi(t)) + \sum_{j \in M}\sum_{i \in N} P^{j,tr}_i(\chi^\xi(t)) + Co]\}\} + \mathbb{E}\{\sum_{j \in M} Q_j(t)[\sum_{i \in N} \alpha^j_i(\chi^\xi(t)) C^j_i(\chi^\xi(t)) - \mu_j(\chi^\xi(t))]\} + \mathbb{E}\{\sum_{i \in N} Z_i(t)[\gamma_i(\chi^\xi(t)) - a_i(\chi^\xi(t))]\} + \mathbb{E}\{\sum_{i \in N} S_i(t)[a_i(\chi^\xi(t)) - \sum_{j \in M} \alpha^j_i(\chi^\xi(t)) C^j_i(\chi^\xi(t))]\} \leq$$

$$\vartheta + V\eta^\xi(t)[\sum_{j\in M}\mathbb{E}\{P_j^c(\chi^\xi(t))\} + \sum_{j\in M}\sum_{i\in N}\mathbb{E}\{P_i^{j,tr}(\chi^\xi(t))\} + Co] - \varepsilon[\sum_{j\in M}Q_j(t) + \sum_{i\in N}S_i(t)] \quad (37)$$

By taking expectations with respect to the distribution of $\Theta(t)$ and using the law of iterated expectations, and summing over $\varsigma \in \{0,1,\ldots,t-1\}$ for some slot $t > 0$, we have

$$\mathbb{E}\{L(\Theta(t)) - L(\Theta(0))\} - V\sum_{\varsigma=0}^{t-1}\mathbb{E}\left\{\begin{array}{c}\sum_{i\in N}U_i(\gamma_i(\varsigma)) - \eta(\varsigma)*\\ [\sum_{j\in M}P_j^c(\varsigma) + \sum_{j\in M}\sum_{i\in N}P_i^{j,tr}(\varsigma) + Co]\end{array}\right\} \le$$
$$\vartheta t + V\sum_{\varsigma=0}^{t-1}\mathbb{E}\{\eta^\xi(\varsigma)\}[\sum_{j\in M}\mathbb{E}\{P_j^c(\chi^\xi(\varsigma))\} + \sum_{j\in M}\sum_{i\in N}\mathbb{E}\{P_i^{j,tr}(\chi^\xi(\varsigma))\} + Co] - \varepsilon\sum_{\varsigma=0}^{t-1}\mathbb{E}\{\sum_{j\in M}Q_j(\varsigma)\} - \varepsilon\sum_{\varsigma=0}^{t-1}\mathbb{E}\{\sum_{i\in N}S_i(\varsigma)\} \quad (38)$$

By dividing both sides of (38) with $t\varepsilon$, using the fact that $\mathbb{E}\{L(\Theta(t))\} > 0$, and arranging the termss, we obtain

$$\frac{1}{t}\sum_{\varsigma=0}^{t-1}\mathbb{E}\{\sum_{j\in M}Q_j(\varsigma)\} + \frac{1}{t}\sum_{\varsigma=0}^{t-1}\mathbb{E}\{\sum_{i\in N}S_i(\varsigma)\} \le \frac{\vartheta}{\varepsilon} + \mathbb{E}\{L(\Theta(0))\} + V\frac{1}{t\varepsilon}\sum_{\varsigma=0}^{t-1}\mathbb{E}\{\eta^\xi(\varsigma)\}[\sum_{j\in M}\mathbb{E}\{P_j^c(\chi^\xi(\varsigma))\} + \sum_{j\in M}\sum_{i\in N}\mathbb{E}\{P_i^{j,tr}(\chi^\xi(\varsigma))\} + Co] + V\{\frac{1}{\varepsilon}\sum_{i\in N}U_i\left(\frac{1}{t}\sum_{\varsigma=0}^{t-1}\mathbb{E}\{\gamma_i(\varsigma)\}\right) - \frac{1}{t\varepsilon}\sum_{\varsigma=0}^{t-1}\mathbb{E}\{\eta(\varsigma)*[\sum_{j\in M}P_j^c(\varsigma) + \sum_{j\in M}\sum_{i\in N}P_i^{j,tr}(\varsigma) + Co]\}\} \quad (39)$$

Then by taking $\lim_{t\to\infty}(.)$, denoting maximum total energy consumption of the system as $e_{max}$ and using Lemma 3, we have

$$\lim_{t\to\infty}\frac{1}{t}\sum_{\varsigma=0}^{t-1}\left(\mathbb{E}\left\{\frac{\sum_{j\in M}Q_j(\varsigma)}{|M|}\right\} + \mathbb{E}\left\{\frac{\sum_{i\in N}S_i(\varsigma)}{|N|}\right\}\right) \le$$
$$\lim_{t\to\infty}\frac{1}{t}\sum_{\varsigma=0}^{t-1}(\mathbb{E}\{\sum_{j\in M}Q_j(\varsigma)\} + \mathbb{E}\{\sum_{i\in N}S_i(\varsigma)\}) \le \frac{\vartheta}{\varepsilon} + V\frac{\eta}{\varepsilon}[\sum_{j\in M}\mathbb{E}\{P_j^c(\chi^\xi(\varsigma))\} + \sum_{j\in M}\sum_{i\in N}\mathbb{E}\{P_i^{j,tr}(\chi^\xi(\varsigma))\} + Co] \le \frac{\vartheta}{\varepsilon} + V\frac{\eta e_{max}}{\varepsilon} \quad (40)$$

And the average mean of queue backlogs bound is proved.

Next, we proceed to prove the bound of utility-power efficiency.

Plugging the task scheduling policy $(\chi^\xi(t), \gamma^\xi)$, which meets the Slater type condition and only depends on i.i.d. system state with stationary distribution, into the right-side of (37), we can deduce

$$\mathbb{E}\{L(\Theta(t+1)) - L(\Theta(t))\} - V\mathbb{E}\left\{\begin{array}{c}\sum_{i\in N}U_i(\gamma_i(t)) - \eta(t)*\\ [\sum_{j\in M}P_j^c(t) + \sum_{j\in M}\sum_{i\in N}P_i^{j,tr}(t) + Co]\end{array}\Big|\Theta(t)\right\} \le$$
$$\vartheta - V\left\{\sum_{i\in N}U_i(\gamma_i^\xi) - \mathbb{E}\{\eta^\xi(t)*[\sum_{j\in M}P_j^c(\chi^\xi(t)) + \sum_{j\in M}\sum_{i\in N}P_i^{j,tr}(\chi^\xi(t)) + Co]\}\right\} \le \vartheta -$$
$$V\eta_{P2}^*\mathbb{E}\{[\sum_{j\in M}P_j^c(\chi^\xi(t)) + \sum_{j\in M}\sum_{i\in N}P_i^{j,tr}(\chi^\xi(t)) + Co]\} + V\eta^\xi(t)\mathbb{E}\{[\sum_{j\in M}P_j^c(\chi^\xi(t)) + \sum_{j\in M}\sum_{i\in N}P_i^{j,tr}(\chi^\xi(t)) + Co]\} \quad (41)$$

By summing (41) over $\varsigma \in \{0, 1, \ldots, t-1\}$, using the law of iterated expectations, dividing it with $tV$, rearranging terms and using Jensen's inequality, and using the fact that $\mathbb{E}\{L(\Theta(t))\} > 0$, also taking the limit as $t \to \infty$, we have

$$\lim_{t\to\infty}\frac{1}{t}\sum_{k=0}^{t-1}\mathbb{E}\{(E_M(k) + E_N(k) + Co)\eta(k)\} - \sum_{i\in N}U_i\left(\lim_{t\to\infty}\frac{1}{t}\sum_{k=0}^{t-1}\mathbb{E}\{\gamma_i(k)\}\right) \le \frac{\vartheta}{V} - \eta_{P2}^*[\sum_{j\in M}\mathbb{E}\{P_j^c(\chi^\xi(t))\} + \sum_{j\in M}\sum_{i\in N}\mathbb{E}\{P_i^{j,tr}(\chi^\xi(t))\} + Co] + \left(\lim_{t\to\infty}\frac{1}{t}\sum_{k=0}^{t-1}\mathbb{E}\{\eta^\xi(k)\}\right)[\sum_{j\in M}\mathbb{E}\{P_j^c(\chi^\xi(t))\} + \sum_{j\in M}\sum_{i\in N}\mathbb{E}\{P_i^{j,tr}(\chi^\xi(t))\} + Co] \quad (42)$$

Using lemma 3, we have:

$$0 \le \frac{\vartheta}{V} - \eta_{P2}^*[\sum_{j\in M}\mathbb{E}\{P_j^c(\chi'(t))\} + \sum_{j\in M}\sum_{i\in N}\mathbb{E}\{P_i^{j,tr}(\chi'(t))\} + Co] + \eta[\sum_{j\in M}\mathbb{E}\{P_j^c(\chi'(t))\} + \sum_{j\in M}\sum_{i\in N}\mathbb{E}\{P_i^{j,tr}(\chi'(t))\} + Co] \quad (43)$$

Thus, we have:

$$\eta \ge \eta_{P2}^* - \frac{\vartheta}{V[\sum_{j\in M}\mathbb{E}\{P_j^c(\chi'(t))\} + \sum_{j\in M}\sum_{i\in N}\mathbb{E}\{P_i^{j,tr}(\chi'(t))\} + Co]} \quad (44)$$

We know that:

$$\sum_{j\in M}\mathbb{E}\{P_j^c(\chi'(t))\} + \sum_{j\in M}\sum_{i\in N}\mathbb{E}\{P_i^{j,tr}(\chi'(t))\} \ge 0 \quad (45)$$

Thus we obtain:

$$\eta \ge \eta_{P2}^* - \frac{\vartheta}{VCo}$$